\documentclass[12pt]{article}
\usepackage{amsmath,amsthm,amscd,amssymb}
\usepackage{amsfonts}
\usepackage{amsmath}
\usepackage{graphicx}
\usepackage[mathscr]{eucal}
\usepackage[english]{babel}
\newtheorem{prop}{Proposition}[section]

\newtheorem{theo}[prop]{Theorem}
\newtheorem{lem}[prop]{Lemma}

\newtheorem{rem}[prop]{Remark}

\def\1{1\!{\rm l}}

\newcommand{\IR}{{\Bbb R}}
\newcommand{\IRd}{{\Bbb R}^d}

\newcommand{\IZd}{{\Bbb Z}^d}
\newcommand{\IP}{{\mathbb P}}
\newcommand{\HD}{H_\omega}
\newcommand{\HN}{\widetilde{H}_{\omega}}

\newcommand{\0}{_\omega}
\newcommand{\om}{(\omega)}
\newcommand{\eh}{\tfrac{1}{2}}
\newcommand{\IE}{\mathbb E}
\newcommand{\oL}{\frac{1}{\,L^{d}\,}}
\newcommand{\Hs}{\widetilde{H}}
\newcommand{\HLD}{\Hs_{\Lambda_L}^{D}(\omega)}

\newcommand{\Qs}{\mathcal{Q}}
\newcommand{\N}{\widetilde{N}}
\newcommand{\no}{\noindent}
\newcommand{\wt}{\widetilde}
\newcommand{\LL}{\Lambda_{L}}
\newcommand{\Et}{\wt{E}}

\begin{document}
\title {Lifshitz tails for
a percolation model\\ in the continuum.}
\date{}
\maketitle
\begin{center}\bf{{W. Kirsch }} \footnote{FernUniversit\"at Hagen, Germany. e-mail: werner.kirsch@fernuni-hagen.de}
and \bf{{H. Najar}} \footnote{D\'epartement de Math\'ematiques
Physiques I.S.M.A.I. Kairouan, Abd Assed Ibn Elfourat  Kairouan
3100. Tunisie.
 }
\end{center}
\begin{abstract}\no In this work we study Lifshitz tails for
Laplacians in a percolation model on $\IRd$. At any lattice point $i$ in $\IRd$
we remove a set $S+i$ with a certain probability $p$.
We consider the Laplacian on the remaining subset of $\IRd$ with either Dirichlet or Neumann
boundary conditions. We prove that the
integrated density of states exhibits Lifshitz behavior at the
bottom of the spectrum when we consider Dirichlet boundary
conditions, while when we consider Neumann boundary conditions , it
exhibits a van Hove behavior.
\end{abstract}
\quad\vspace{20mm}

\noindent{\small\sf 2000 Mathematics Subject Classification:15A52, 35P05,
37A30,
47F05.}

\newpage
\section{Introduction}
In this paper we study the integrated density of states in the context of a quantum percolation in the continuum.
The Hamiltonian we consider is the Laplacian on a random subset $D_\omega$ of $\IRd$. The set $D_\omega$ is constructed as follows:\\
At any lattice point $i\in\IZd$ we remove from $\IRd$ with probability $p$ and independently from the other lattice points a set around $i$, more precisely we remove the set $S+i=\{x+i\mid x\in S\}$
where $S$ is a compact subset of $\IRd$. $D_{\omega}$ is the what remains of
$\IRd$ after removing these copies of $S$. Let us denote by  $H_{\omega}$  the Dirichlet-Laplacian and by $\Hs_{\omega}$  the Neumann-Laplacian on the set $D_\omega$ respectively.

We will be interested in the integrated density of states $N_{D}$ resp. $N_{N}$ of these operators and in particular their behavior
near the bottom of the spectrum. It turns out that under suitable conditions on $S$ and/or $p$ the density of states shows Lifshitz
behavior in the Dirichlet case, but doesn't in the Neumann case.

The integrated density of states (IDS)
measures the number of energy levels per unit volume, below a given
energy, more precisely: Let $P_{(-\infty,
E]}$ be the spectral projection of a random Schr\"odinger operator $
H_{\omega}$, $\Lambda_{L}$ be a cube in $\IRd$ of side length $L$ around
the origin and $\chi_{A}$ the characteristic function of the set $A\subset\IRd$.  We consider
\begin{equation}
N(E)=\lim_{L\to
\infty}\frac{1}{\mid\Lambda_L\mid}tr(\chi_{\Lambda_L}P(-\infty,E]).\label{equ1}
\end{equation}

\noindent Under quite general assumption, this limit exists and is non random. This is in particular true for our
model operators $\HD$ and $\HN$.
The quantity $N$ called the \emph{integrated density of states} of $%
H_{\omega}$. See \cite{KM} and references given there for an overview on the IDS.

The question we are interested
in here concerns the behavior of $N$ at the bottom of the spectrum of
$H_{\omega}$.
In 1964, Lifshitz \cite{Lif} argued that,
for a Schr\"{o}dinger operator of the form ${H_{\omega}=-\Delta+
V_{\omega}, }$ there exists $c_{1},c_{2}>0$ such that $N(E)$
satisfies the asymptotic:
\begin{equation}
N(E)\simeq c_1\exp({-c_2(E-E_{0})^{-\frac{d}{2}}}),\ \ E\searrow E_{0}.
\label{eq2}
\end{equation}
\noindent Here $E_0$ is the bottom of the spectrum of $H_{\omega}$. The
behavior (\ref{eq2}) is known as \emph{Lifshitz tails}. In the
last thirty years, there has been vast literature, both physical and
mathematical, concerning Lifshitz tails and related phenomena.
 We do not try to give an exhaustive
account of this literature. The paper \cite{KM}
gives a survey of such results and basic references on this subject.
Below, we give results on the IDS behavior in the context of our
percolation operators $\HD$ and $\HN$.

A quantum percolation Hamiltonian was studied already
by de Gennes et al in \cite{Gen1, Gen2}, where the Hamiltonian of
binary solid solution was considered. It is proved that the spectrum
of these percolation Hamiltonians is pure point if the fraction $p$
is less than the critical value $p_c$. We recall that $p_c$ is the
value of the well-known critical probability of the percolation
theory: If $p<p_c$, no infinite active cluster exists almost surely,
and for $p>p_c$ there exists almost surely one infinite cluster.
Theses facts are given in \cite{Gri} and were mainly obtained by
Hammersley in the late fifties. We notice that uniqueness of the
infinite cluster, was proved only thirty years later by Aizenman,
Kesten and Newmann, see \cite{Gri}.

If the concentration of
active sites is above the critical value, one speaks of the \emph{
percolation regime}. For this regime it is argued \cite{Gen1} that
the spectrum contains a continuous part. In \cite{KiMu}, it is
proved that in the non-percolation case, $p\in ]0,p_c[$, the
spectrum of the Laplacian is $\mathbb{P}$-almost surely only a dense
pure-point spectrum with infinitely degenerate eigenvalues.

Bond-percolation graphs are random subgraphs of the $d$-dimensional
integer lattice generated by a standard bond-percolation. The
associated graph Laplacians, subject to Dirichlet or Neumann
conditions at these cluster boundary, represent bounded,
self-adjoint, ergodic random operators with an off-diagonal
disorder. They have almost surely a non-random spectrum.

In
\cite{Cha} the authors considered the site dilution model on the
hyper-cubic lattice $\mathbb{Z}^d$, for $d\geq 2$. They investigated
the density of states for the tight-binding Hamiltonian projected
onto an infinite cluster. It is shown that, almost surely, the IDS
is discontinuous on a set of energies which is dense in the band.
This is proved by constructing states  supported on finite
regions of the infinite cluster.

In the same context, in
\cite{Ves1}, Veseli\'c studied Hamiltonians (a finite hopping range
operators) corresponding to site percolation on the lattice
$\mathbb{Z}^d$ and graphs with an amenable group action and
characterize the set of energies which are almost surely eigenvalues
with infinitely supported eigenfunctions. It is proved that this set
of energies is a dense subset of algebraic integers and this set of
energies corresponds to the discontinuity point of the IDS.

Spectral theory of random graphs, however, is still a widely open
field. The recent contributions \cite{Bar, Hei, Math} take a
probabilistic point of view to derive heat-kernel estimates for
Laplacians on supercritical Bernoulli bond-percolation graphs in the
$d$-dimensional hyper-cubic lattice. On the other hand, spectral
theory methods are used by Kirsch and M\"uller in \cite{KiMu} to
study spectral properties of the Laplacian on bond-percolation
graphs. Indeed they investigate the IDS of Laplacian  on subcritical
bond-percolation graphs. Depending on the boundary condition that is
chosen at cluster borders, two different types of Lifshitz
asymptotics at spectral edges were proved, precisely at the lower
spectral edge for bond probabilities $p<p_c$, the IDS, $\N(E)$ of
the Neumann Laplacian satisfies
\begin{equation}
\lim_{E \to 0^{+}} \frac{\log|\log \big( \N(E)-\N(0)
\big)|}{\log E}= -\frac{1}{2}.\label{eq3}
\end{equation}
Here one notices that the Lifshitz exponent $\frac{1}{2}$ in
(\ref{eq3}) is independent of the spatial dimension $d$. (see
equation  \ref{eq2}, for the usual dependence). This is due to the
fact that asymptotically, $\N$ is dominated by the smallest
eigenvalues which are caused by very long linear clusters. In
contrast, for $p<p_c$, it is
proved that the integrated density of states $N(E)$ of the
Dirichlet Laplacian satisfies
\begin{equation}
\lim_{E \to 0^{+}} \frac{\log|\log \big( N(E) \big)|}{\log E}=
-\frac{d}{2}.\label{eq4}
\end{equation}
In (\ref{eq4}), the Lifshitz exponent is in the classical form i.e
is $\frac{d}{2}$, this is explained by the fact that this is, the
dominating small Dirichlet  eigenvalues arise from large fully
connected cube-or sphere-like clusters. We notice that due to some
symmetries the Lifshitz tails at the upper spectral edge are related
to the ones at the lower spectral edge whereas at the upper spectral
edge, the behavior is reversed.\newline For the dual case to
\cite{KiMu}, M\"uller and Stollmann in \cite{MuSt} pursue the
investigation of \cite{KiMu} and studied spectral asymptotics of the
Laplacian on supercritical ($p<p_c$) bond-percolation graphs. They
studied the influence and the contribution of the existence of the
infinite cluster. The situation is different. Indeed, in the present
situation it is proved that $N_N$ exhibits van Hove asymptotics.
Precisely
\begin{equation}
\lim_{E \to 0^{+}} \frac{\log \big( \N(E)-\N(0)%
\big)}{\log E}= \frac{d}{2}.\label{eq5}
\end{equation}
We notice that (\ref{eq5}) is due to the existence of the
$d$-dimensional infinite grid.  In contrast to the Neumann case, for the
Dirichlet Laplacian (\ref{eq4}) is still true for $p\geq
p_c$ situation.

Lifshitz tails for Neumann Laplacian on Erd\"os-R\'enyi
random graphs at the lower spectral edge $E=0$, are considered in
\cite{KhKiMu}.

\section{Model and results}
\subsection{The Quantum Percolation Hamiltonian}
We start by describing our percolation Hamiltonian.
Let $S_{0}\subset\IRd$ be a bounded open set and denote by $S$ the
closure of $S_{0}$. Furthermore, let  $\{\omega_\gamma\}_{\gamma\in\IZd}$  be a sequence of independent random variables with
$\IP(\omega_\gamma=1)=p$ and $\IP(\omega_\gamma=0)=1-p$ and set $\Xi_\omega=\{\gamma\mid \omega_\gamma=1\}$.
Then we define the random sets
\begin{equation}
\Gamma_\omega=\bigcup_{i\in\Xi_\omega} (S+i)\qquad\text{and}\qquad D_\omega=\IRd\setminus\Gamma_\omega
\end{equation}

\noindent Finally we denote by $H_{\omega}$ and $\Hs_{\omega}$ the Laplacian on $D_\omega$ with Dirichlet resp. Neumann boundary conditions at $\partial D_{\omega}$.

We will always follow the convention to decorate quantities related to \textit{Neumann
boundary conditions} at $\partial D_{\omega}$ by a tilde, for example the corresponding
operator is denoted by $\Hs_{\omega}$, its integrated density of states by $\N(E)$ and so
on. The quantities for \textit{Dirichlet boundary conditions} will not be decorated, i.~e.
will be denoted by $H\0$, $N(E)$ etc.

For the operators $H\0$ and $\Hs\0$ we will also have to consider the restrictions
to $\Lambda_{L}=[-\tfrac{L}{2},\tfrac{L}{2}]^{d}$. We denote the restriction of $H\0$
to $\Lambda_{L}$ with Dirichlet and Neumann conditions at $\partial\Lambda_{L}$
by $H_{L}^{D}\om$ and $H_{L}^{N}\om$ respectively. Similarly, $\Hs_{L}^{D}\om$ and
$\Hs_{L}^{D}\om$ denote the restriction of $\Hs\0$ to $\Lambda_{L}$ with Dirichlet resp.
Neumann boundary conditions at the boundary of the cube $\Lambda_{L}$. So, for
example, the operator $\Hs_{L}^{D}$ is defined on $L^{2}(\Lambda_{L}\cap D\0)$
and has Neumann boundary conditions on $\partial D\0\cap\Lambda_{L}$ and
Dirichlet boundary conditions on $\partial \Lambda_{L}\setminus D\0$. Finally we
will take the liberty to supress the argument $\omega$ when the dependence of
a quantity on $\omega$ is clear from the context.

In an informal way we may write $H\0=-\Delta +V_{\omega}$,
with the random `potential':

\begin{eqnarray}\label{informal1}
V_{\omega}&=&\sum_{\gamma\in \mathbb{Z}^d}\omega_{\gamma}\,f(x-\gamma)\\[3mm]
 \ \ \text{with\qquad}\ \ f(x)&=&\begin{cases}\; \infty, &{\text{for }}x\in\mathcal{S}\\
\; 0 &
 \ {\text{elsewhere }}\end{cases}\label{informal2}
\end{eqnarray}

If the set $S$ is contained in $]-\tfrac{1}{2},\tfrac{1}{2}[^d$ then the set $D\0$ contains a unique unbounded cluster independent of
the value of $p$. In fact, $D\0$ always contains the set
\begin{equation}\label{D1}
D_1=\IRd\setminus\bigcup_{i\in\IZd}(S+i).
\end{equation}
Following our general convention, we denote by $H_{1}$ the Laplacian on $D_{1}$
with Dirichlet boundary conditions and by $\Hs_{1}$ the same operator with Neumann boundary conditions.

Whenever $D_1$ contains an unbounded component
then $D\0$ will as well. On the other hand if $D_1$ contains no unbounded cluster then $D\0$ may or may not contain an unbounded cluster depending on the value
of $p$ and the shape of the set $S$. If $D\0$ contains only bounded components
then both $H\0$ and $\Hs\0$ have pure point spectra.
If not stated otherwise we always assume from now on that $S\subset]-\tfrac{1}{2},\tfrac{1}{2}[^d$.

%%\textbf{Attention: Rewrite paragraph\\ include: measurability (problem: domains),
% order inequality in proof (2.8), spectrum for p=1!}

The families $H\0$ and $\Hs\0$ are ergodic families of self-adjoint operators,
more precisely:
%Let us consider the
%map  $\mathcal{P}$, from $\Omega$, to the set of the self-adjoint
%operator on $L^2(\mathbb{R}^d)$, such that to $\omega$ associate
%$\displaystyle \mathcal{P}(\omega)=H_{\omega}=
%H_{\omega}^{\bullet}=-\Delta^{\bullet}_{\lceil
%\mathbb{R}^d\backslash \Gamma_{\omega}}, \bullet\in\{N,D\}$.
%$\mathcal{P}$ is measurable.
\newline Let $\mathcal{U}_{i}$ be the
unitary translation operator on $L^2(\mathbb{R}^d)$ given by
$$\mathcal{U}_{i}\psi (x)=\psi (x-i),\ \ \
\forall \psi \in L^2(\mathbb{R}^d) \ \text{and}\ \
x\in\mathbb{R}^d.$$ As the probability measure $\mathbb{P}$ is
ergodic with respect to the group of translation
$(\mathcal{T}_{i})_{i\in \mathbb{Z}^d}$, acting as $\displaystyle
\mathcal{T}_{i}(\omega)= (\omega_{\gamma+i})_{\gamma \in
\mathbb{Z}^d}$, we get
\begin{equation}
\mathcal{T}_{i}^{-1}H_{\omega}\mathcal{T}_{i}=H_{(\mathcal{T}_i\omega)},\
\ \forall i\in \mathbb{Z}^d, \ \omega\in \Omega.
\end{equation}
%By this, we deduce that $H_{\omega}$ is a measurable family of
%self-adjoint and ergodic operators.
We may therefor apply the methods from \cite{kir1, KiMaC,PaFi} to
conclude that there exists $\Sigma, \Sigma _ {pp}, \Sigma _ {ac}$ and
$\Sigma _ {sc}$ closed and non-random sets of ${\Bbb{R}}$ such that
$\Sigma$ is the spectrum of $H_{\omega}$ with probability one and
such that if $\sigma_{pp}$ (respectively $\sigma_{ac}$ and
$\sigma_{sc}$) denote the pure point spectrum (respectively the
absolutely continuous and singular continuous spectrum) of
$H_{\omega}$, then $\Sigma _ {pp}=\sigma _ {pp}, \Sigma _
{ac}=\sigma _ {ac}$ and $\Sigma _ {sc}=\sigma _ {sc}$ with
probability one.

There is a little subtlety connected with the question of
measurability in our case. Since the Hilbert spaces the operators
$H\0$ and $\Hs\0$ act on \emph{depend} on $\omega$ we need
a notion of measurability appropriate for our situation. We get around this
problem by noticing that the kernel of the operators $e^{-tH\0}$ and
$e^{-t\Hs\0}$ can be expressed via the Feynman-Kac formula (see for
example \cite{KiMa} and references given there). We extend the operators
$e^{-tH\0}$ and $e^{-t\Hs\0}$ to $L^{2}(\IRd)$ by extending it by the zero
operator on $L^{2}(\IRd\setminus D\0)$. The corresponding operators are
easily seen to be measurable in the sense of \cite{KiMaC}, as the kernels
are explicitly measurable. Moreover these operators form ergodic
families. Consequently their spectra and the above defined
parts of the spectra are non random. Thus, the same assertion is true
for $H\0$ and $\Hs\0$ as their spectra can be computed from the spectra of
 $e^{-tH\0}$ and $e^{-t\Hs\0}$.

 The following Lemma gives the precise
location of the spectrum.

\begin{lem}\label{lem1}
Suppose that $S\subset]-\tfrac{1}{2},\tfrac{1}{2}[^{d}$.
 \begin{enumerate}
 \item For $0\leq p<1$, the spectrum $\Sigma$, of $H_{\omega}$ is $[0,+\infty[$ with
probability one.
\item
 For $0\leq p\leq 1$, the spectrum $\widetilde{\Sigma}$, of $\Hs_{\omega}$
 is $[0,+\infty[$ with probability one.
\end{enumerate}
\end{lem}
\begin{rem}
Unless $S$ is empty the infimum of the spectrum of $H_{1}$ will be strictly
positive, so in part 1 of the above lemma we had to exclude the case $p=1$.
In contrast to this the spectrum of $\Hs_{1}$ always contains $0$. In fact the constant function on $D_{1}$ (or on $D\0$ in general) is a generalized eigenfunction for $E=0$.
\end{rem}
\noindent{\bf \textit{Proof:}} First let us notice that for any $\omega \in
\Omega$, we have
\begin{equation} \label{equ2.9}
H\0~\ge 0 \textnormal{\quad and\quad} \Hs\0~\ge~0
\end{equation}
so
\begin{equation}\label{spos}
\Sigma~\subset~[0,\infty[ \textnormal{\quad and\quad}
\widetilde{\Sigma}~\subset~[0,\infty[
\end{equation}

\noindent To complete the proof we have to show the opposite
inclusion, i.e
\begin{equation}[0,+\infty[ \subset \Sigma\   \text{for}\
\mathbb{P}-{\text{almost\  every}} \omega \in \Omega. \label{equ7}
\end{equation}

\no We do this for $H\0$, the proof for $\Hs\0$ is essentially the same.

\no For this, let $\overline{\Omega}$, be the following events
\begin{eqnarray}
\overline{\Omega}&=&\Big\{ \omega \in \Omega\mid\ \ \text{For any}\  n\in \mathbb{N},
\text{there exists a }\notag\\\quad  && x_{n}\,\big(=x_{n}\om\big)\in\IZd \text{ such that }
\big(x_{n}+\Lambda_{n+1}\big)\,\cap D\0~=~\emptyset\Big\}
\label{equ8}
\end{eqnarray}
Let $E\in
[0,+\infty[=\Sigma(-\Delta)$ be arbitrarily fixed.
 Using Weyl criterion, we know that there exists a Weyl sequence
 $(\varphi_{E,n})_{n\in \mathbb{N}}\subset L^2(\mathbb{R}^d)$
 , for $-\Delta$. Thus $\|\varphi_{E,n}\|=1,$ for all $n\in \mathbb{N}$ and
\begin{equation}
 \lim_{n\to \infty} \|(\Delta+E\cdot  \mathbb{I})\varphi_{E,n}\|=0
 \end{equation}
 Notice that for any $i\in \mathbb{Z}^d$, $(\mathcal{U}_i\varphi_{E,n})_{n\in \mathbb{N}}$ is also a Weyl sequence.
 Without loss of generality, we assume that the sequence $(\varphi_{E,n})_{n\in \mathbb{N}}$ is compactly supported.
 So for any $\omega \in \overline{\Omega}$, there exists a Weyl sequences $(\varphi_{E,n}^{\omega})_{n\in \mathbb{N}}$ for
  $(-\Delta)$ such that
\begin{equation}
\textrm{supp}\,\varphi_{E,n}^{\omega}\subset x_{n}+\Lambda_{n}
\end{equation}
with $x_{n}$ as in \eqref{equ8}. By definition of $\overline{\Omega}$ we
have
\begin{equation}
\textrm{supp}\,\varphi_{E,n}^{\omega}\;\cap\;D\0~=~\emptyset
\end{equation}
Consequently
for any $n\in \mathbb{N}$ and $\omega \in \widetilde{\Omega}$,
$\varphi_{E,n}^{\omega}$ is in the domain of the operator $H\0$ and we get
\begin{equation}
\|(H_{\omega}-E\mathbb{I})\varphi_{E,n}^{\omega})\|=\|(\Delta+E\cdot
\mathbb{I})\varphi_{E,n}^{\omega}\|.
   \end{equation}
Hence, $(\varphi_{E,n}^{\omega})_{n\in \mathbb{N}}$ is also a Weyl
sequence for $H_{\omega}$. So we get (\ref{equ7}) for any $\omega
\in \overline{\Omega}$.

It remains to check that
$\mathbb{P}(\overline{\Omega})=1$. Define
\begin{eqnarray}
\overline{\Omega}_{n}&=&\Big\{ \omega \in \Omega\mid\ \ \text{there exists a
sequence }  y_{k}\in\IZd \notag\\ &&\text{ such that for all } k\quad
\big(y_{k}+\Lambda_{n+1}\big)\,\cap D\0~=~\emptyset\Big\}.
\end{eqnarray}

By the Borel-Cantelli lemma we know that $\IP\big(\overline{\Omega}_{n}\big)=1$.
Since $\overline{\Omega}\supset\bigcap\overline{\Omega}_{n}$ we conclude
that $\mathbb{P}(\overline{\Omega})=1$.
 $\hfill \Box$

 \subsection{The main results}
We investigate the integrated densities of states $N_{D}(E)$ and $N_{N}(E)$
for energies $E$ near $0$, the bottom of the almost sure spectrum of both
$H_{\omega}^{D}$ and
 $H_{\omega}^{N}$. The first
 result says that in the Dirichlet case our Percolation Hamiltonian has a
 \textsl{Lifshitz singularity} there, as one might guess from \eqref{informal1} and \eqref{informal2}:
 \begin{theo}\label{theo1} Assume $S\subset]-\eh,\eh[^{d}$ and $p\in ]0,1[$,
 then
 \begin{equation}
 \lim _{E\to 0^+}\frac{\log|\log N_D(E)|}{\log E}=-\frac{d}{2}.
 \end{equation}
 \end{theo}
 \vspace{7mm}
 At a first glance one might be tempted to expect the same behavior for
 $N_{N}$. However, this is not the case. It turns out, that $0$ is a stable
 boundary for the Neumann operator in the sense of \cite{PaFi} and we
 have a \textsl{van Hove singularity} in that case:
 \begin{theo}\label{theo2}  Assume $S\subset]-\eh,\eh[^{d}$ and $p\in[0,1]$,
 then there is a constant $C>0$ such that for small $E>0$
 \begin{equation} N_{N}(E)~\ge~C\,E^{d/2}.\label{equ2.14}\end{equation}
 \end{theo}
If
 the set $S$ has a `hole' in the sense that the complement $\complement S$
 of $S$ contains a bounded connected components $M$ then the constant function
 on $i+M$ is an eigenstate to energy $E=0$ whenever $\omega_{i}=1$.
 Thus $0$ is an eigenvalue of $\Hs_{\LL}^{X}$ for $X=N, D$ whose multiplicity
 is proportional to $\text{vol}(\LL)$ for typical $\omega$. Hence, the integrated
 density of states will be discontinuous at $0$ whenever $\complement S$ is not
 connected and $E=0$ is an eigenvalue of $\Hs\0$ with infinite multiplicity.

 On the other hand, if $S$ has no `holes', i.\,e. if $\complement S$ is connected, then
 $0$ is not an eigenvalue of $\Hs\0$ almost surely.

\section{Proofs}
\subsection{Preliminaries}\label{sec3}
We start by recalling the
following result and giving some properties of the IDS.
\begin{prop}
Let $\varphi \in C_{0}^{\infty}(\mathbb{R}^d)$, then
\begin{equation}
\lim_{L\to \infty}\frac{1}{\text{vol}(\Lambda_L)}
tr(\varphi(H_{\omega})\chi_{\Lambda_L})=\mathbb{E}\Big(tr(\chi_{\Lambda_1}
\varphi(H_{\omega})\chi_{\Lambda_1})\Big),\label{equ11}
\end{equation}
for $\mathbb{P}$-almost all $\omega$. Here $\mathbb{E}$, is the
expectation with respect to the probability measure $\mathbb{P}$.
\end{prop}
{\textit{\bf{Proof:}}} First we write $\Lambda_L=\sum_{i\in
\Lambda_{L}\cap \mathbb{Z}^d}\Lambda_1(i)$, Here $\Lambda_1(i)$, is
the cube of center $i$ end side length $1$. We set
$\zeta_i=tr(\varphi(H_{\omega})\chi_{\Lambda_1(i)}). $ So $\zeta_i$
is an ergodic sequence (with respect to $\mathbb{Z}^d$) of random
variables. So \begin{equation}\frac{1}{\text{vol}(\Lambda_L)}
tr(\varphi(H_{\omega})\chi_{\Lambda_L})=\frac{1}{\text{vol}(\Lambda_L)}\sum_{i\in
\Lambda_{L}\cup \mathbb{Z}^d}\zeta_i. \label{equ12}
\end{equation}
By the Birkhoff's ergodic theorem, the sum in (\ref{equ12})
converges to its expectation value. This ends the proof of
(\ref{equ11}).$\hfill\Box$\newline Now, we notice that both sides of
(\ref{equ11}), are positive linear functionals on the bounded,
continuous functions. So, they define positives measures
respectively $\mu_{L}$ and $\mu$.
i.e$$\int_{\mathbb{R}}\varphi(\lambda)d\mu_{L}
(\lambda)=\frac{1}{\text{vol}(\Lambda_L)}
tr(\varphi(H_{\omega})\chi_{\Lambda_L})$$ and
$$\int_{\mathbb{R}}\varphi(\lambda)d\mu
(\lambda)=\mathbb{E}\Big(tr(\chi_{\Lambda_1}
\varphi(H_{\omega})\chi_{\Lambda_1})\Big).$$ For those two measures
we have the following result proved in \cite{Ves1},
\begin{theo}For almost all $\omega\in \Omega $ and for all $\varphi\in
C_{0}^{\infty}({\Bbb{R}})$  we
have
$$\lim_{k\rightarrow \infty
}\langle \varphi,d\mu_{L}\rangle=\langle \varphi,d\mu\rangle.$$
\end{theo}
\begin{rem}
We call the non-random probability measure $\mu$ {\textit{the
density of states measure}}. It verifies the following
fundamental properties
$$N(E)=\mu((-\infty,E]),$$
$$
\Sigma(H_{\omega})={\textrm{supp}(\mu)}.
$$
\end{rem}
As mentioned above we denote the operator $H\0$ restricted to $L^{2}(\Lambda_{L})$
with Dirichlet boundary conditions at $\partial\LL$ by $H_{\LL}^{D}=H_{\LL}^{D}\om$,
and the corresponding operator with Neumann boundary conditions at
$\partial\LL$ by $H_{\LL}^{N}=H_{\LL}^{N}\om$. We use an analogous notation
for $\Hs\0$.

Since these operators have compact resolvents their spectra are purely discrete.
We order their eigenvalues in increasing order with repetition of eigenvalues
according to multiplicity. The eigenvalues of $H_{\LL}^{X}$ with $X=D$ or $X=N$
are denoted by
  \begin{equation}
  E_1^{X}(\Lambda_L)\leq  E_2^{X}(\Lambda_L)\leq \cdots
  \leq  E_n^{X}(\Lambda_L)\leq \cdots
  \end{equation}
  and those of $\Hs_{\LL}^{X}$ by
  \begin{equation}
  \Et_1^{X}(\Lambda_L)\leq  \Et_2^{X}(\Lambda_L)\leq \cdots
  \leq  \Et_n^{X}(\Lambda_L)\leq \cdots
  \end{equation}

 If $A$ is any operator with discrete spectrum, bounded below
 and $E\in\IR$ define $N(A,E)$ to be the
 number of eigenvalues of $A$ less than or equal to $E$, of course counted with their
  multiplicities.

To prove Theorem \ref{theo1}, we prove a lower and an upper bounds on $N(E)$.
 The upper and lower bounds are proven separately and based on the following
 result (see \cite{KiMa} or \cite{PaFi}).
 \begin{equation}
 \frac{1}{\mid\Lambda_L \mid}\mathbb{E}\{N(H_{\Lambda_L}^D(\omega),E)\}
 \leq N(E)\leq \frac{1}{\mid\Lambda_L \mid} \mathbb{E}\{N(H_{\Lambda_L}^N(\omega),E)\}. \label{exe1}
 \end{equation}
 and
  \begin{equation}
 \frac{1}{\mid\Lambda_L \mid}\mathbb{E}\{N(\Hs_{\Lambda_L}^D(\omega),E)\}
 \leq \wt{N}(E)\leq \frac{1}{\mid\Lambda_L \mid} \mathbb{E}\{N(\Hs_{\Lambda_L}^N(\omega),E)\}. \label{exe2}
 \end{equation}
 Inequalities in (\ref{exe1}) and \eqref{exe2} are based on the method of Neumann-Dirichlet bracketing (see \cite{ReSi} and \cite{KiMa}). Indeed
 \begin{equation}
 H_{\Lambda_1}^N(\omega)\oplus H_{\Lambda_2}^N\leq H_{\Lambda_1 \cup
 \Lambda_2}^N (\omega)
 \end{equation}
 and
 \begin{equation}
H_{\Lambda_1 \cup
 \Lambda_2}^D(\omega) \leq  H_{\Lambda_1}^D(\omega)\oplus
 H_{\Lambda_2}^D(\omega),
 \end{equation}
 hold on $L^2(\Lambda_1 \cup \Lambda_2),$ for all bounded cubes $\Lambda_1,\Lambda_2\subset \mathbb{R}^d$
 whenever the interior of $\Lambda_1 \cap \Lambda_2$ is empty (\cite{ReSi}).
 \subsection{The Dirichlet case:}
\subsubsection{The upper bound} The upper bound
 is proved by a comparison procedure. Indeed,
 let,
$$\widetilde{V}_{\omega}=\sum_{\gamma \in
\mathbb{Z}^d}\omega_{\gamma} \chi_{\mathcal{S}}(x-\gamma), $$ and
$$\widetilde{H}_{\omega}=-\Delta +\widetilde{V}_{\omega}$$
For any $\Lambda_L\subset \mathbb{R}^d$, we set \begin{equation}
\mathcal{Q}^{\Lambda_L}_{\omega}(\varphi,\psi)=\langle\varphi, H_0
\psi\rangle,\ \ \ \varphi,\psi \in H_0^1(\Lambda_L\backslash
\Gamma_{\omega})=\mathcal{D}_{\Lambda_L},
\end{equation}
and
\begin{equation}
{\widetilde{\mathcal{Q}}^{\Lambda_L}}_{\omega}(\varphi,\psi)=\langle\varphi,
\widetilde{H}_{\omega} \psi\rangle,\ \ \ \varphi,\psi \in
H_0^1(\Lambda_L)=\widetilde{\mathcal{D}}_{\Lambda_L}.
\end{equation}
From \cite{ReSi}, we recall the following result,
\begin{lem}\label{ha1}
 For any $L,k\in \mathbb{N}^*$, we have
\begin{multline}\sup_{\varphi_1,\cdots, \varphi_{k-1}\in
\widetilde{\mathcal{D}}_{\Lambda_L}}\inf_{\psi\in
[\varphi_1,\cdots,\varphi_{k-1}]^{\bot}\cap
{\widetilde{\mathcal{D}}}_{\Lambda_L} ,
\|\psi\|=1}{\widetilde{\mathcal{Q}}}^{\Lambda_L}_{\omega}(\psi,\psi)\leq
\\ \sup_{\psi_1,\cdots,\psi_{k-1}\in\mathcal{D}_{\Lambda_L}}\inf_{\varphi\in
[\psi_1,\cdots,\psi_{k-1}]^{\bot}\cap\mathcal{D}_{\Lambda_L}
,\|\varphi\|=1}\mathcal{Q}^{\Lambda_L}_{\omega}(\varphi,\varphi) .
\end{multline}
\end{lem}
From Lemma \ref{ha1}, one deduces that for any $n\in \mathbb{N}^*$,
we have \begin{equation}E_n(\widetilde{H}_{\omega}(\Lambda_L))\leq
E_n(H_{\omega}(\Lambda_L))  .\end{equation} Thus, we get that for
any $E\in \mathbb{R}$,
\begin{equation} N(H_{\Lambda_L}(\omega),E)\leq
N(\widetilde{H}_{\Lambda_L}(\omega),E)
\end{equation} We notice that for
$\widetilde{H}_{\omega}$ it is already known that it exhibits
Lifshitz tails, by the result of Kirsch and Simon \cite{KirSim2}.
This ends the proof of the upper bound. $\hfill \Box$
\subsubsection{The lower bound} We recall that for any
$\Lambda_L\subset \mathbb{R}^d$, we have
\begin{equation}
H_\omega \leq H_{\Lambda_L}^D(\omega) \leq H_{1,\Lambda_L}^D.
\end{equation}
So by the min-max argument we get that
\begin{equation}
E_{1}^D(\Lambda_L)\leq E_{1}^D(H_{1,\Lambda_L}).
\end{equation}
 Using equation (\ref{exe1}) one gets,
\begin{eqnarray}
N(E)&\geq&\frac{1}{L^d}\cdot
\mathbb{P}\{E_1^D(\Lambda_L)\leq  E\}\nonumber \\
&\geq & \frac{1}{L^d}\cdot \mathbb{P}\{E_{1}^D(\Lambda_{L})
\leq E \ \text{and}\  \forall \gamma\in \Lambda_L\cap \mathbb{Z}^d,\ \omega_\gamma=0\} \nonumber \\
&= &\frac{1}{L^d}\cdot  \mathbb{P}\{E_1^D(H_{0,\Lambda_L})\leq E \
\text{and}\ \forall \gamma\in \Lambda_L\cap \mathbb{Z}^d,\
\omega_\gamma=1\}\nonumber\\  &\geq&
\mathbb{P}\{\omega_0=0\}^{|\Lambda_L|}=(1-p)^{L^{d}}\label{equ}.
\end{eqnarray}
By this, we deal with the estimate of the volume of $\Lambda_L$ i.e
the order of $L$. As $H_{0,\Lambda_L}$ is the free Laplacian
restricted to $\Lambda_L$, it is known that
$E_1^D(-\Delta_{\Lambda_L})\simeq\frac{1}{L^2}$. So to be less than
$E$, $L$ should be $c\cdot E^{-\frac{d}{2}}$. This ends the proof of
the lower bound.

\subsection{The Neumann case}
We estimate:
\begin{eqnarray}
N_{N}(E)~&\ge&~\oL\,\IE\Big(N\big(\Hs_{\Lambda_L}^{D}(\omega),E\big)\Big)\notag\\
&\ge&~\oL\,\IP\Big(E_{1}\big(\Hs_{\Lambda_{L}}^{D}(\omega)\big)\le E\Big)\label{estN}
\end{eqnarray}
for arbitrary $L$. By the min-max-principle we have
for any $\psi\in \Qs\big(\HLD\big)$, $\psi\not=0$, we have
\begin{equation*}
E_{1}\big(\HLD\big) ~\leq~
\frac{\langle \nabla\psi, \nabla\psi\rangle_{D_{\omega}^{L}}}
{\langle \psi,\psi \rangle_{D_{\omega}^{L}}}
\end{equation*}
where we used $\langle\cdot,\cdot\rangle_{D_{\omega}^{L}}$ to denote the scalar product
in the space $L^{2}(D_{\omega}^{L})=L^{2}(D_{\omega}\cap\Lambda_{L})$. We conclude
\begin{equation}\label{test}
N_{N}(E)~\ge~\oL\,\IP\Big(\langle \nabla\psi, \nabla\psi\rangle_{D_{\omega}^{L}}
~\le~E\,\langle \psi,\psi \rangle_{D_{\omega}^{L}}\Big)
\end{equation}
Now we construct a test function $\psi\,(=\phi_{L})$ as follows:
Let $\phi$ be a smooth function on $\IR^{d}$ with $\textrm{supp}\,\phi\subset ]-\eh,\eh[^{d}$,
$0\le \phi(x)\le 1$ and $\phi(x)=1$ for $x\in[-\tfrac{1}{4},\tfrac{1}{4}]^{d}$.
We set:
\begin{equation}
\phi_{L}(x)~=~\frac{1}{L^{d/2}}\;\phi\big(\frac{x}{L}\big)
\end{equation}
It follows that $\phi_{L}$ (or rather its restriction to $D_{\omega}^{L}$) belongs to $\Qs(\HLD)$ for all $\omega$, so we may
take $\phi_{L}$ as a test function in \eqref{test}.

We have
\begin{eqnarray}
\langle \nabla\phi_{L}, \nabla\phi_{L}\rangle_{D_{\omega}^{L}}~&\le&
\oL\,\int_{\Lambda_{L}}\;\frac{1}{L^{2}}\,\big|(\nabla\phi)\big(\frac{x}{L}\big)\big|^{2}\;dx
\notag\\[2mm]
&\le&~C_{1}\,\frac{1}{L^{2}}
\end{eqnarray}
and
\begin{eqnarray}
\langle \phi_{L},\phi_{L}\rangle_{D_{\omega}^{L}}~&\ge&~
\oL\;\int_{D_{1}^{L}}\,\big|\phi(\frac{x}{L}\big|^{2}\;dx\notag\\[2mm]\notag
&\ge&~\oL\;\textrm{vol}(D_{1}\cap\Lambda_{L/4})\\[2mm]
&\ge&~C_{2}
\end{eqnarray}
where $C_{2}>0$ is a constant which only depends on the volume $\textrm{vol}(M)$
of the set $M$.

Thus we have proved that $N_{N}(E)\ge\oL$ as longs as $C_{1}\tfrac{1}{L^{2}}\le
C_{2}E$. This can be guaranteed by choosing $L=C'\,E^{-1/2}$ with a suitable constant
$C'$. Thus we proved:

\begin{equation}
N_{N}(E)~\ge~C\,E^{d/2}
\end{equation}
for some constant $C>0$.

\textit{$\mathbf{Acknowledgements.}$ This work was made possible through visits
of H.\,N. in Germany and of W.\,K. in Tunisia. Both authors are grateful for
the warm hospitality and financial support of the Univerity of Kairouan,
the FernUniversit\"at Hagen and the SFB/TR 12 of the Deutsche Forschungsgemeinschaft. }

\end{document}